\documentclass[reprint, amsmath,amssymb, aps,showkeys,prd]{revtex4-2}

\usepackage{amsmath}
\usepackage{amssymb}
\usepackage{amsfonts}
\usepackage{graphicx}
\usepackage{dcolumn}
\usepackage{bm}
\usepackage{color}
\usepackage{theorem}
\usepackage{enumerate}
\usepackage[utf8]{inputenc}
\usepackage{setspace}
\usepackage{titlesec}

\titleformat{\paragraph}
{\normalfont\normalsize\bfseries}{\theparagraph}{1em}{}
\titlespacing*{\paragraph}
{0pt}{3.25ex plus 1ex minus .2ex}{1.5ex plus .2ex}

\newcommand{\metric}{\ensuremath{\mathrm{g}}}
\newcommand{\submetric}{\ensuremath{\mathrm{h}}}
\newcommand{\normal}{\ensuremath{\mathrm{u}}}

\newcommand{\hubble}{\ensuremath{\mathrm{H}}}

\newcommand{\grad}{\raisebox{0.01em}{\scalebox{0.9}[1.3]{${\scriptstyle \nabla}$}}}

\newcommand{\anisotropy}{\raisebox{0.01em}{\scalebox{0.9}[1.3]{${\scriptstyle \Sigma}$}}}
\newcommand{\anisotropym}{\raisebox{0.01em}{\scalebox{0.8}[0.8]{${\mathcal{N}}$}}}

\newcommand{\pressure}{\ensuremath{\mathrm{P}}}
\newcommand{\energy}{\ensuremath{\mathrm{\rho}}}

\newcommand{\curvature}{\ensuremath{\mathrm{R}}}
\newcommand{\state}{\textrm{w}}

\newcommand{\deviation}{\scalebox{0.7}[0.8]{\ensuremath{\Sigma}}}

\begin{document}
%


%
\title{Spatially flat universes with isotropic tidal forces}

\author{Fabio Scalco Dias}
 \email{scalco@unifei.edu.br}
\author{Leandro Gustavo Gomes}
 \email{lggomes@unifei.edu.br}
\author{Luis Fernando  Mello}
 \email{lfmelo@unifei.edu.br}
 \affiliation{Universidade Federal de Itajub\'a, Av. BPS, 1303,  Itajub\'a-MG, Brazil. ZIP: 37500-903 .
}

\date{\today}

\begin{abstract}
We investigate the dynamics of the spatially flat universes submitted to isotropic tidal forces and adiabatic expansion under Einstein's equations. Surprisingly, the tendency to a high Hubble anisotropy at late times starts to appear as far as we assume the strong energy condition, a characteristic which becomes dominant in the radiation era and even more stringent under a stiff matter regime. We introduce the parameter $b$ which measures the relative change in the magnitudes of the Hubble anisotropy and the scale factor and use it to show that the anisotropies must be kept small as long as we assume the Universe has passed through an inflationary period, in accordance with the cosmic no-hair theorem. Hence, we have a class of models that shows us in a simple and straightforward way the instability of the FLRW universes, the furtiveness of the isotropy concept, and how they can still be consistent with the standard model of Cosmology as far as we assume the occurrence of an inflationary early period.   
\end{abstract}

\maketitle

\section{Introduction}

Isotropy is a quite elusive property in Cosmology. The Hubble ratio of expansion vary no more than $1 \%$ along the different directions of the sky \cite{Soltis_2019,Tedesco_2019}, the CMB temperature is known to be independent of direction with a high degree of precision \cite{Saadeh_PRL, Isotropy_CMB_2020}, and the number count of radio sources seems to be consistent with the isotropy hypothesis \cite{Bengaly_2019}. Therefore, as we apply the Copernican principle, meaning that we are not privileged observers of the cosmic drift, we would be in the comfortable position to set our spacetime as a small perturbation of a homogeneous and isotropic FLRW universe, the current standard model of Cosmology \cite{Weinberg_2008}. That would be just as perfect as one could expect, except for one important point: the FLRW universes are not stable under small perturbations. 

Historically, just after the discovery of the CMB in the '60s, which by the time was showing its first traces of isotropy, the picture of a homogeneous, chaotic, and highly anisotropic early epoch had been proposed \cite{BKL82}, often referred to as the BKL scenario. The current state of the universe would be achieved as the anisotropy dies out during the expansion, which could be caused by neutrino viscosity, for instance \cite{Misner67,Misner68}. Soon after that, Collins and Hawking showed that the spatially homogeneous universes do not, in general, isotropize \cite{CollinsHawking1973}. That was a distinguished point in the conceptual evidence for the instability of the FLRW models, which in turn became a barrier for the chaotic BKL picture \cite{Barrow82}. Since then, the BKL approach survived as a general framework for approaching the big bang singularity \cite{Berger2014}, the dynamical aspects of the anisotropies have been further studied \cite{LeBlanc_1997,Calogero2008, BSGC_2021,DSGM_2022}, and the inflationary theory has taken into the scene, leaving no cosmological hair to account neither for homogeneity nor isotropy \cite{Wald1983}. 

Today, some decades after the results of Collins and Hawking, the behavior of the anisotropy under Einstein's dynamics can still baffle our intuition, even in the simplest of the models, under the most reasonable physical conditions. In order to show that, we start with a general spatially flat model, a Bianchi type I spacetime, which is put in an adiabatic expansion with only one restriction: the gravitational forces are everywhere isotropic, so that the gravitational pull (or push) felt by the components of our free-falling cosmic fluid has no distinguished direction. As we are going to show, even under this framework favoring an isotropic behavior, the tendency to anisotropy still persists, and when the thermodynamic pressure is as high as in the case of pure radiation, it dominates completely. They form a class of simple and rather counter-intuitive examples, which has a three-fold implication for Cosmology: (i) It enhances our conceptual understanding of the cosmic anisotropies. In particular, it gives us a reminder of how the FLRW universes are unstable under small perturbations, even when the overall situation is favoring stability; (ii) It displays a complete nonlinear account for the anisotropy dynamics in a quite simple and physically reliable manner. This could be useful in case the observations start pointing to an anisotropic Hubble sky \cite{Schucker_2014, Zhao_2022}, which would demand models beyond the small perturbations of an isotropic background; (iii) Even if the observed universe comes to be plainly isotropic, it also furnishes a good class of examples where the cosmic no-hair theorem can be appreciated and clearly understood. As we are going to see, during an inflationary period, the Hubble anisotropy in our model decays twice as fast as the scale factor. Hence, we are able to follow straightforwardly how inflation would drive the leftovers of the cosmic anisotropies to an insignificant position in the cosmic evolution.

The manuscript is divided as follows: in the section \ref{Sec:IsotropyTidalForces}, we define what we mean by isotropy in the gravitational tidal forces and adiabatic expansion in the spatially flat models. In the following section, we describe the Einstein's equations in this setting, showing the first traces of the growth of the anisotropies. In the section \ref{Sec:AnisotropyDynamics}, we specialize the dynamics to those cases with a linear equation of state between the energy density and the thermodynamic pressure. This simplification gives us a whole class of examples where the global anisotropy dynamics can be seen in the Kasner disc, enlightening our understanding of such behaviors. In particular, we introduce a parameter $b$ which measures the relative change in the magnitudes of the Hubble anisotropy and the scale factor. This has an important role in justifying why the anisotropies must be kept small after an inflationary period. At the final section, we make our concluding considerations. The notations and sign conventions follow the reference \cite{MTW}.

\section{Isotropy of the tidal forces and adiabatic expansion in spatially flat universes}\label{Sec:IsotropyTidalForces}

Our main assumption is that the universe allows a class of free-falling observers to whom the space sections are flat and the tidal forces are isotropic. This means that the gravitational forces should balance out in a way that the pull (or push) felt by any of these observers is the same in any direction. As we put this forward in the mathematical language, the observers are represented by a unitary, geodesic, and vorticity-free fundamental velocity, $\normal$, whose space sections are flat, that is, a Bianchi type I model, in a way the tidal force operator, $-\, \curvature^\mu_{\kappa \nu \lambda} \normal^\kappa \normal^\lambda$, has no preferred spatial directions. This means that it is multiple of the identity along the spatial directions, which is equivalent to the equation of state \cite{ellis_mac_marteens}
\begin{equation}\label{Eq:TidalIsotropyEqState}
E^\mu_\nu = \tfrac{1}{2}\, \Pi^\mu_\nu \, ,
\end{equation}
where $E^\mu_\nu$ is the electric part of the Weyl tensor and $\Pi^\mu_\nu$ the anisotropic stress tensor. In this case, as we use adapted coordinates $(t,x^i)$ for which $\normal=\partial_t$ and the metric is
\begin{equation}
\metric = -\, dt^2 + \submetric_{ij}(t)\, dx^i\, dx^j \, ,
\end{equation}
the geodesic deviation equations along the spatial directions turn into
\begin{equation}
\frac{d^2\xi^i}{d \tau^2}
= -\, \frac{1}{6}\, \left(\energy + 3 \, \pressure \right)\, \xi^i \, ,
\end{equation}
where $\energy = T_{\mu\nu}\normal^{\mu}\normal^{\nu}$ and $\pressure = T_i^i$  stand for the energy density and total pressure, respectively. The Hubble and the matter-radiation anisotropies will be identified with the dimensionless spatial tensors \cite{DSGM_2022} $\anisotropy_i^j= \sigma_i^j/(\sqrt{6}\,\hubble)$ and $\anisotropym_i^j= \Pi_i^j/(\sqrt{6}\, \pressure)$, respectively, where $\sigma_i^j$ is the shear tensor. Hence, the electric Weyl tensor becomes
\begin{equation}
\frac{1}{\sqrt{6}}\, E_k^i
=
\hubble^2 \, \anisotropy_k^i
-\, \frac{1}{2}\, \pressure\, \anisotropym_k^i
- \,  \sqrt{6} \, \hubble^2 \, \left(
\anisotropy^i_\ell\anisotropy^\ell_k - \, \frac{\anisotropy^2}{3}\, \delta^i_k
\right) \, .
\end{equation}
Finally, we can re-write the equation of state (\ref{Eq:TidalIsotropyEqState}) as
\begin{equation}\label{Eq:TidalIsotropyEqState2}
\pressure\, \anisotropym^i_k
=
\hubble^2\, \left( \anisotropy_k^i
- \,  \sqrt{6} \, \left(
\anisotropy^i_\ell\anisotropy^\ell_k - \, \tfrac{\anisotropy^2}{3}\, \delta^i_k
\right)  \right) \, .
\end{equation}

We assume an adiabatic expansion, with no energy flux, $q^i=0$, nor any spatial heat diffusion, $\partial_iT=0$, so that the entropy is conserved: $\dot{S}=0$. As we split the pressure in its bulk and thermodynamic counterparts, $\pressure_B$ and $\pressure_T$, respectively, we obtain \cite{ellis_mac_marteens}
\begin{equation}
\frac{d S}{d t}=0 \quad \Rightarrow \quad
\pressure_B = \pressure - \pressure_T =
-\, 2\, \pressure\, \anisotropym_{ik}\, \anisotropy^{ik} \,.
\end{equation}
This means that the bulk pressure is composed by the Hubble and matter-radiation components of the anisotropy. On the other hand, the energy conservation, $\grad_\mu \,T^\mu_0=0$, turns out to be dependent only on the thermodynamical pressure, since it is written as
\begin{equation}\label{Eq:ConservationEnergy}
\frac{d \energy}{d t}
= -\, 3  \hubble \left( \energy + \pressure_T \right) \, .
\end{equation}
It is worth mentioning that this equation emulates the conservation of energy in the FLRW spacetimes. Therefore, $\pressure_T$ should be interpreted as the ``isotropic'' part of the total pressure, which is responsible for the net force the cosmic fluid exerts upon the fabric of the spacetime. On the other hand, $\pressure_B$ accommodates that part remaining in the process of taking thermodynamical averages, no less important, since they are connected to the entropy increase.

\section{The cosmic dynamics from Einstein's equations}

The Einstein's equations in the variables $\anisotropy_k^i$ and $\anisotropym_k^i$ have been put forward in Ref. \cite{DSGM_2022}. They  are equivalent to the Generalized Friedmann equation,
\begin{equation}\label{Eq:Friedmann}
\energy = 3\, \left(\, 1- \anisotropy^2\, \right)\, \hubble^2  \, ,
\end{equation}
where $\anisotropy=\sqrt{\anisotropy_k^i\anisotropy^k_i}$ is the Hubble anisotropy magnitude, the conservation of energy (\ref{Eq:ConservationEnergy}) and the anisotropy equation, which after applying the condition (\ref{Eq:TidalIsotropyEqState2}) for isotropic tides, becomes 
\begin{equation}\label{Eq:AnisotropyEquation}
\frac{1}{\hubble}\frac{d}{dt}\, \anisotropy_k^i
= \left( 1  - \frac{\energy -\pressure}{2\, \hubble^2} \right)\, \anisotropy_k^i
- \,  \sqrt{6}  \, \left(
\anisotropy^i_\ell\anisotropy^\ell_k - \, \frac{\anisotropy^2}{3}\, \delta^i_k \right) \, .
\end{equation}

The first consequence of the tidal anisotropy is that the relation (\ref{Eq:TidalIsotropyEqState2}) allows $\anisotropy^i_k$ and $\anisotropym^i_k$ to be simultaneously diagonalizable, and hence, as we put $\anisotropy^i_k(t_0)$ in the diagonal form, the Einstein's equations tell us that it will continue to be like that along the entire expansion. In other words, our Bianchi I spacetime is diagonalizable. Hence, the system (\ref{Eq:AnisotropyEquation}) is completely determined by the equations for the Hubble anisotropy magnitude $\anisotropy$ and the Kasner angle $\alpha$, as 
\begin{equation}
(\anisotropy^k_i)=\sqrt{\frac{2}{3}}\,\anisotropy\,
\left[
\begin{array}{ccc}
 \sin\alpha & 0 & 0 \\
0 & \sin\left( \alpha+\frac{2}{3}\pi\right) & 0\\
0 & 0 & \sin\left( \alpha-\frac{2}{3}\pi\right) \\
\end{array}
\right] \, .
\end{equation}
From this and the relation (\ref{Eq:TidalIsotropyEqState2}), we can write the bulk pressure as
\begin{equation}
\pressure_B =  -\, 2\, \anisotropy^2\,\hubble^2\, \left(1+ \anisotropy\, \sin(3\alpha) \, \right) \, .
\end{equation}
By using the new time parameter $ds=\hubble dt$ and the ``equation-of-state" variable $\state$, that is, 
\begin{equation}\label{Eq:EquationOfStateVariable}
s(t)=\ln \left(\frac{a(t)}{a_0}\right) \quad \textrm{and} \quad \state(t) = \frac{\pressure_T(t)}{\energy(t)} \, ,
\end{equation}
the energy conservation becomes
\begin{equation}\label{Eq:EinsteinEquationsConservation}
\energy'=-3 (\state+1)\, \energy \, ,
\end{equation}
while the anisotropy equation turns into
\begin{equation}\label{Eq:EinsteinEquationsKasnerDisc}
\left\{
\begin{aligned}
\anisotropy'
&=
\anisotropy\, \left( 1-\anisotropy^2 \right)\,
\left( \anisotropy\, \sin(3 \alpha) + \, \frac{3 \state -1}{2} \right),
\\
\alpha'
&=
\anisotropy \, \cos (3 \alpha),
\end{aligned}
\right.
\end{equation}
where we have used the abreviation $z'=dz/ds$. We will consider only the inner part of the Kasner disc ($\anisotropy \le 1$), since this is equivalent of keeping the energy density non-negative, according to the generalized Friedmann equation (\ref{Eq:Friedmann}).

\section{The cosmic dynamics for fluids with a linear equation of state}
\label{Sec:AnisotropyDynamics}

In order to have a glimpse of the different features of the dynamical behavior of our cosmic system, we will consider the expansion with $\state$ constant, that is, with the energy density and the thermodynamical pressure satisfying the linear equation of state $\energy = \state\, \pressure_T$, with $\state'=0$.

\subsection{General properties of the solutions}

Equations (\ref{Eq:EinsteinEquationsKasnerDisc}) define a smooth and autonomous system in the Kasner disc $\anisotropy \le 1$. In the ``Cartesian" coordinates, $x=\deviation\,\cos(3\alpha)$ and $ y=\deviation\, \sin(3\alpha)$, it turns out to be polynomial, as
\begin{equation}\label{Eq:EinsteinEquationsKasnerDiscCartesian}
\left\{
\begin{aligned}
x' & = x\left(1 - x^2 - y^2\right)\left(y + \frac{3\state - 1}{2}\right)  - 3xy, \\
y' & = y\left(1 - x^2 - y^2\right)\left(y + \frac{3\state - 1}{2}\right)  + 3x^2. \\
\end{aligned}
\right.
\end{equation}
Since all the solutions are kept inside the compact disc, they are defined for every real value of $s$, that is, for every $a > 0$. Hence, for all of them, as we assume expansion ($\dot{a}>0$), we have two distinct epochs, just as in the FLRW case: the early ($a \ll a_0$) and the late-time ($a\gg a_0$) universes. Furthermore, the conservation (\ref{Eq:EinsteinEquationsConservation}) is also analogous to its counterpart in the isotropic universes, so that the energy density turns out to be
\begin{equation}
 \energy(t) = \energy_0\, \left(\frac{a_0}{a(t)}\right)^{3(\state+1)} \, .
\end{equation}
The anisotropy magnitude $\anisotropy$ and the Kasner angle $\alpha$, in general, cannot be fully integrated from the equations in (\ref{Eq:EinsteinEquationsKasnerDisc}). Notwithstanding, as we observe that
\begin{equation}
\frac{d\anisotropy}{\anisotropy (1-\anisotropy^2)}
=
\tan (3\alpha) d\alpha + \frac{3 \state -1}{2}\frac{da}{a}
\end{equation}
whenever $\cos(3\alpha_0) \ne 0$, we obtain the constraint
\begin{equation}\label{Eq:SolutionConstraint}
\anisotropy(t) = \frac{\anisotropy_0}{\sqrt{\anisotropy_0^2 + (1-\anisotropy_0^2) \xi(t)}}
\end{equation}
with
\begin{equation}
\xi(t) = \left(\frac{a_0}{a(t)}\right)^{3 \state -1}
\left(\frac{\cos (3 \alpha(t))}{\cos (3 \alpha_0)}\right)^{\frac{2}{3}} \, .
\end{equation}
On the other hand, the solutions with $\sin(3 \alpha_0)=\pm 1$ satisfy $\alpha'=0$. Hence, by a direct integration of the first of the equations in (\ref{Eq:EinsteinEquationsKasnerDisc}), we obtain
\begin{equation}\label{Eq:SolutionSinConstant}
\int_{\anisotropy_0}^{\anisotropy(t)}
\frac{d\anisotropy}{\anisotropy\, \left( 1-\anisotropy^2 \right)\,
\left(  \frac{3 \state -1}{2} \pm \anisotropy \right)}
=
\ln \left(\frac{a(t)}{a_0}\right) \, .
\end{equation}
Note that in the generic case (\ref{Eq:SolutionConstraint}), when $\state > 1/3$, as we set $a \to \infty$ we get $\xi \to 0$ and $\anisotropy \to 1$. This implies that these solutions get more and more anisotropic as the universe expands. This rather counter-intuitive behavior, since the tides are kept isotropic, persists even when $\state$ attains smaller values, up to the breaking point of the strong energy condition, $\state =- 1/3$. This important fact will be addressed throughout the text.

It is interesting to analyze the relation of the orders of magnitude both the scale factor and the anisotropy went through between two specific moments of the expansion of the universe, say from $t_1$ to $t_2$. This is characterized by the parameter
\begin{equation}\label{Eq:RatioParameter}
b = \left|\frac{\ln(\anisotropy_2/\anisotropy_1)}{\ln(a_2/a_1)}\right| \, .
\end{equation}
As we analyze it along the solutions $\sin(3 \alpha_0)=\pm 1$, we have, according to (\ref{Eq:SolutionSinConstant}),
\begin{equation}\label{Eq:RatioParameterIntegral}
\frac{1}{b} =
\left|
\frac{1}{\ln(\anisotropy_2/\anisotropy_1)}
\int_{\anisotropy_1}^{\anisotropy_2}
\frac{d\anisotropy}{\anisotropy\, \left( 1-\anisotropy^2 \right)\,
\left( \anisotropy \pm \frac{3 \state -1}{2} \right)}
\right|\, .
\end{equation}
As we take $\anisotropy_2/\anisotropy_1 \to 0$ or $\anisotropy_1/\anisotropy_2 \to 0$ in (\ref{Eq:RatioParameterIntegral}), we conclude that 
\begin{equation}\label{Eq:RatioParameterLimit}
b \approx  \frac{|3 \state -1|}{2} 
\end{equation}
whenever one of the variables $\anisotropy_1$ or $\anisotropy_2$ overcomes the other in many orders of magnitude. These formulas will be of suitable usage in order to estimate the variation of the anisotropy magnitudes during the different epochs of the universe.

\subsection{The qualitative aspects of the dynamics}

The equilibrium points of the system (\ref{Eq:EinsteinEquationsKasnerDisc}), with $\state$ constant, are the origin, $\anisotropy =0$, representing the flat FLRW universe, the Taub points in the Kasner circle $\anisotropy=1$, and the LRS points inside the disc \footnote{For the Taub points, see \cite{BSGC_2021}. Any point with Kasner angular coordinates $n\pi/6$, $n$ odd, is Locally Rotationally Symmetric (LRS), which refers to the more symmetric configuration of the spacetime. A good analogy is to compare the ellipsoids of revolution in the Euclidean spaces (LRS) with their less symmetric partners.}, with $\anisotropy = |3 \state -1|/2$. They come in two categories: the $T$'s and the $Q$'s. The Taub points $T_1, T_2, T_3$, with $\anisotropy =1$, and the LRS ones $\widetilde{T}_1,\widetilde{T}_2,\widetilde{T}_3$, with $\anisotropy = (3 \state -1)/2$, these last ones existing only in the case $1/3 \le \state \leq 1$, have the following Kasner angles coordinates, respectively,  
\begin{equation*}
\alpha_1 = \frac{\pi}{2}, \,
\alpha_2 = \frac{11 \pi}{6}, \,
\alpha_3 = \frac{7 \pi}{6}. \\
\end{equation*}
The Taub points $Q_1, Q_2, Q_3$, with $\anisotropy =1$, and the LRS ones $\widetilde{Q}_1,\widetilde{Q}_2,\widetilde{Q}_3$, with $\anisotropy = (1-3 \state)/2$, these last ones existing only in the case $-1 \le \state \leq 1/3$, have the following Kasner angles coordinates, respectively,
\begin{equation*}
\alpha_1 = \frac{3\pi}{2}, \,
\alpha_2 = \frac{5 \pi}{6}, \,
\alpha_3 = \frac{\pi}{6}.
\end{equation*}
Note that when $\state=-1/3$, the points $\widetilde{Q}_i$'s  coincide with the $Q_i$'s, and when $\state=1$, the points $\widetilde{T}_i$'s coincide with the $T_i$'s. Moreover,
as $\state \to 1/3^{+}$,  $\widetilde{T}_i$ approach the origin, as well as $\widetilde{Q}_i$, when $\state \to 1/3^{-}$.

The linear part of system \eqref{Eq:EinsteinEquationsKasnerDiscCartesian} at the origin is
\begin{equation}\label{Eq:LinearOrigin}
\left[ \begin {array}{cc} (3\state-1)/2 &0
\\\noalign{\medskip}0& (3\state-1)/2 \end {array} \right].
\end{equation}
At the Taub points, the linear part of system (\ref{Eq:EinsteinEquationsKasnerDisc}) is
\begin{equation}\label{Eq:LinearPartKasnerDynamicsP}
\left[ \begin {array}{cc} 3-3\state &0
\\\noalign{\medskip}0&3\end {array} \right]  \quad \\  \quad \text{or} \quad
\left[ \begin {array}{cc} -(3\state+1) &0
\\\noalign{\medskip}0&-3\end {array} \right] \, ,
\end{equation}
whether the point is $T_i$ or $Q_i$ $(i=1, 2, 3)$, respectively. Finally, for the points $\widetilde{T}$'s or $\widetilde{Q}$'s, we obtain
\begin{equation}\label{Eq:LinearPartKasnerDynamicsQ}
\left[ \begin {array}{cc} \frac{3}{8}(9\state^2-1) (\state-1) &0
\\\noalign{\medskip}0&\frac{3}{2}(3\state - 1))\end {array} \right].
\end{equation}
In short, the stability of each of these points is given in Table \ref{Tab1} where the following notation is used: S (saddle), UN (unstable node), SN (stable node)  HS (hyperbolic sectors) and S-N (saddle-node) \cite{DLA_2007}.

\begin{table}[!ht]
\centering
{\small
{\begin{tabular}{|c|c|c|c|c|c|c|c|c|}
\hline
{$\state$ / Point} & Origin     &$Q_i$  & $T_i$  & $\widetilde{Q}_i$  & $\widetilde{T}_i$            \\ \hline
$\state \in [-1,-1/3)$              & SN   & S      & UN      & $\nexists$     & $\nexists$   \\  \hline
$\state= -1/3$                       & SN   & S-N   & UN      & $Q_i$  & $\nexists$    \\  \hline
$\state \in (-1/3,1/3)$            & SN   &SN     & UN      & S        & $\nexists$   \\  \hline
$\state= 1/3$                        & 6 HS     &SN      & UN      & Origin  & Origin \\  \hline
$\state \in (1/3,1)$                & UN   &SN       & UN      & $\nexists$  & S \\  \hline
$\state= 1$                           & UN   &SN       & S-N     & $\nexists$  & $T_i$  \\  \hline
\end{tabular} }
\caption{Topological type of each equilibrium point of system (\ref{Eq:EinsteinEquationsKasnerDisc}) for $-1\leq \state\leq 1$. }
\label{Tab1}
}
\end{table}
In what follows, we analyze the dynamics in the Kasner disc ($\anisotropy \le 1$) with the variables $\anisotropy$ and $\alpha$ working as ``polar coordinates'' in the plane \cite{BSGC_2021,DSGM_2022}. We are interested in the interval $-1 \le \state \le 1$.  We could abuse of our intuition and refer to the condition $\state=0$ as ``dust'', $\state=1/3$ as ``radiation'', $\state=1$ as ``stiff-matter'', and so on. This would be justified as far as the anisotropies are kept small, so that we could interpret those situations as small perturbations of the proposed physical situation. In the case the anisotropies grow large, those proposed names could be quite misleading. For this reason, we will adopt the names ``dust-like'', ``radiation-like'', ``stiff-matter-like'', and so on.

\subsubsection{From the inflation-like scenario to the breaking point of the strong energy condition}

\begin{figure}[h!]
\centerline{
\includegraphics[width=5cm]{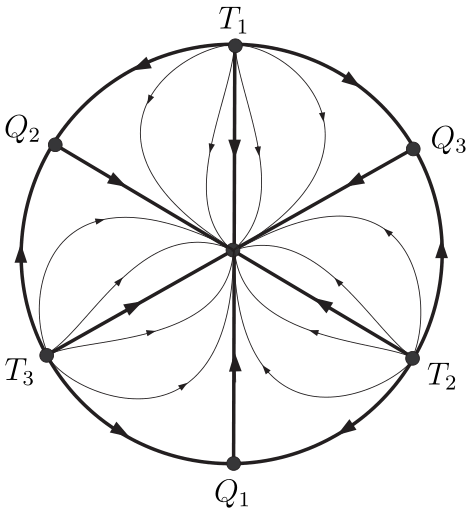}}
\vspace{-0.4cm}
\caption{\small The phase portrait in the Kasner disc ($\anisotropy \le 1$) for $-1 \le \state \le -1/3$. Any solution isotropizes towards the FLRW universe at the center. The $T$-Taub points are unstable nodes. The $Q$-Taub points are saddles up to the value $\state = -1/3$, when they coincide with the $\widetilde{Q}$'s and begin the transition to  stable nodes. In this case, they are saddle-nodes.}
\label{Fig1}
\end{figure}

According to the standard picture of Cosmology, the interval $ -1 \leq \state < -1/3$ encompass the very early inflationary era as well as the late-time dark energy period, both satisfying $\pressure_T \approx -\energy$. The state $\state = -1/3$ will be referred to as the breaking point where the strong energy condition \footnote{Indeed, the strong energy condition demands $\energy + 3 \pressure >0$ and $\energy + \pressure + \Pi_i >0$, $i=1,2,3$, $\Pi_i$ the eigenvalues of $\Pi_{ij}$. The name refers to the state at which it begins to be violated, at least for small anisotropy ($\anisotropy \approx 0$).}. Here, the $T$-Taub points are unstable nodes while the $Q$'s are saddles, with all the solutions inside the Kasner disc converging to the late-time isotropic cosmology, that is, $\anisotropy \to 0$ as $a \to \infty$ (see Fig. \ref{Fig1}).

Since the anisotropy decreases as the universe expands, let us pick an initial and a final states, $\anisotropy_1$ and $\anisotropy_2$, respectively, with $\anisotropy_2 \ll \anisotropy_1$. According to the formula (\ref{Eq:RatioParameterLimit}), we have $b \approx (1-3\state)/2$. This means that the anisotropy diminishes twice as fast as the universe expands, if $\state =-1$, or at the same ratio, if $\state = -1/3$. Let us analyze the first case separately, due to its conceptual importance. 

The parameter $b$ in the case $\state = -1$ can be straightforwardly calculated through the integral (\ref{Eq:RatioParameterIntegral}) along the solution $\alpha = \pi/2$. If we take $\anisotropy_2=\anisotropy_1\times 10^{-n}$, $n>1$, we obtain 
\begin{equation}
\frac{2}{b}-1\approx \frac{-\ln(1-\anisotropy_1)+\ln(1+\anisotropy_1/2)-2\ln(1+\anisotropy_1)}{3 n \ln 10} \, ,
\end{equation}
where we have used $\ln(1+ \anisotropy_1\times 10^{-n}) \approx 0$. For $\anisotropy_1$ not too close to $1$, we have $b\approx 2$, just as aforementioned. On the other hand, if the initial condition was extremely anisotropic, as $\anisotropy_1=1-10^{-n_1}$, we would have $b \approx 2- 2 n_1/(3 n +n_1)$, meaning that $b$ could attain smaller values, but we would still have $b \ge 4/3$. Hence, when $\pressure_T = - \energy$, the anisotropy vanishes faster than the universe expands. Furthermore, if the universe have passed through $N$ e-folds during this period, the anisotropy would diminish something near to $2N$ e-orders of magnitude. This is in agreement with the no-hair picture of the cosmic evolution \cite{Wald1983}. As we apply it to the inflationary period, where the universe is believed to stay long enough as $N \gtrsim 60$ \cite{ellis_mac_marteens}, we would have the anisotropy at the beginning of the reheating period probably as tiny as $\anisotropy_{rh} \lesssim e^{-120}$. If not that, at least it would not be greater than $e^{-80}$, as we put $b=4/3$. An analogous situation would occur during the late-time dominance of the dark energy, but now with an inferior value for $N$.

\subsubsection{From the breaking point of the strong energy condition to the radiation-like condition}

From the dynamical viewpoint, the interval $-1/3 \le \state < 1/3$  is characterized by the origin still being a stable node, but at this time its basin of attraction is not the inner Kasner disc anymore. In fact, new saddle-type equilibrium points appear along the straight segments connecting the $Q$-Taub points to the origin, those LRS ones labeled as $\widetilde{Q}_1$, $\widetilde{Q}_2$ and $\widetilde{Q}_3$. To each of them, there correspond two separatrices splitting the disc into two parts: the inner one, which contains the basin of attraction for the equilibrium at the origin, where the universe tends to a late-time isotropic state, and the outermost one, where the anisotropization takes place, and the solutions tend to the highly anisotropic LRS universes at the $Q$-Taub points.
\begin{figure}[h!]
\centerline{
\includegraphics[width=5cm]{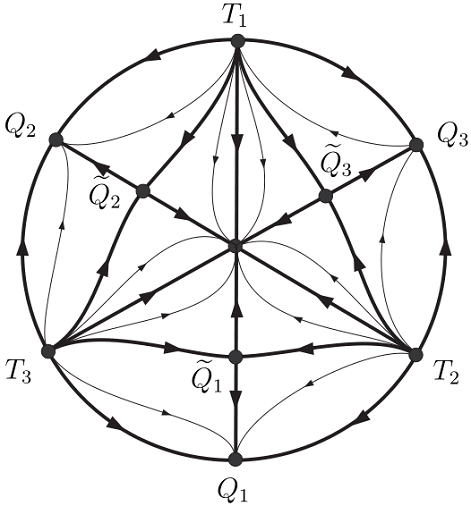}}
\vspace{-0.4cm}
\caption{\small The phase portrait in the Kasner disc ($\anisotropy \le 1$) for $-1/3 < \state < 1/3$. There are separatrices sppliting the dynamics in two main global behaviors. In the innermost part, the solutions isotropize towards the FLRW universe at late-times. In the outermost one, the universes tend to the highly anisotropic LRS models at the $Q$-Taub points. As we get closer to the radiation epoch ($\state \to 1/3^{-}$), anisotropization starts to dominate over isotropization.}
\label{Fig2}
\end{figure}

The $\widetilde{Q}$'s equilibrium points at $\anisotropy = (1-3\state)/2$ split the line $\alpha = (1+4n)\pi/6$ in two: the isotropization and anisotropization segments, where $0 <\anisotropy <(1-3\state)/2$ and $(1-3\state)/2 < \anisotropy <1$, respectively (see Fig. \ref{Fig2}). As we estimate the order of magnitude parameter $b$ with the aid of the formula (\ref{Eq:RatioParameterLimit}), we obtain
\begin{equation}
\anisotropy_2 \ll \anisotropy_1 < \frac{1-3\state}{2}
\quad \Rightarrow \quad
b= \frac{1-3\state}{2} \, .
\end{equation}
Therefore, as $\anisotropy_1 < \frac{1-3\state}{2}$, the anisotropy decays slower than the universe expands, by a factor $b$, with $0< b < 1$, such that $b \to 1$ as $\state \to -1/3^{+}$ and $b \to 0$ as $\state \to 1/3^{-}$. In the case of dust-like solutions ($\state =0$), we have that each two e-folds of the universe corresponds to one of the anisotropy, such that $b=1/2$, as long as $\anisotropy_1<1/2$. On the other hand, when $\anisotropy_1 > \frac{1-3\state}{2}$, the parameter $b$ loses its general character, so that the estimate it is intended for should be directly calculated from the integral (\ref{Eq:RatioParameterIntegral}), since it can give any positive number.

\subsubsection{The radiation-like era}

\begin{figure}[h!]
\centerline{
\includegraphics[width=5cm]{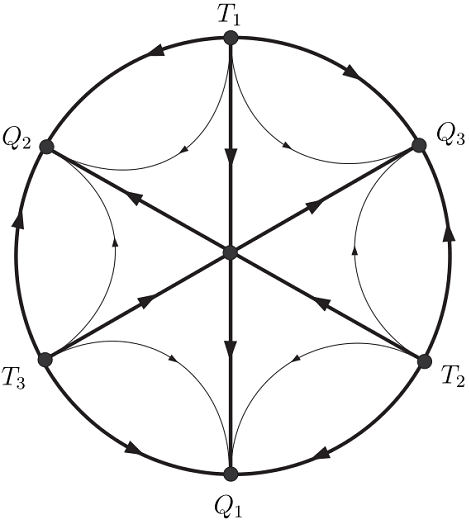}}
\vspace{-0.4cm}
\caption{\small  The phase portrait in the Kasner disc ($\anisotropy \le 1$) for the radiation era ($\state = 1/3$). The solutions anisotropize towards the LRS models at the $Q$-Taub points, where $\anisotropy=1$, except the FLRW universe at the origin and those at the straight segment with $\alpha=\pi/2+2n\pi/3$.  }
\label{Fig3}
\end{figure}

The radiation-like era is distinguished as being the transition point to pure anisotropization. In fact, the $\widetilde{Q}$-type equilibrium points coalesce with the origin, so that the anisotropization sector becomes virtually the only one available inside the Kasner disc, except for the segments where $\sin(3\alpha_0) = -1$, the remains of the former isotropization region. Therefore, the isotropic universe is no longer stable for small perturbations, for $\anisotropy \to 1$ as $a \to \infty$, no matter the initial values as far as we keep $\anisotropy_0 \ne 0$ and $\sin(3\alpha_0) \ne -1$. The special cases where $\sin(3\alpha_0) = -1$ are still tending to the late-time FLRW model. The dynamics in the Kasner disc for this epoch is depicted in Fig. \ref{Fig3}.

In the case $\cos(3\alpha_0) \ne 0$, we obtain from the formula (\ref{Eq:SolutionConstraint}) the first integral
\begin{equation}\label{Eq:SolutionConstraintRadiation}
\frac{(1-\anisotropy^2)^{3/2}}{\anisotropy^3 \cos(3\alpha)} =c_0.
\end{equation}
This implies that system (\ref{Eq:EinsteinEquationsKasnerDisc}) is integrable for $\state =1/3$.

Let us assume that our model describes the hot and dense epoch dominated by radiation, when the input anisotropy was $\anisotropy_1$, and by the end of this period, when the CMB was released from the initial plasma and started to propagate freely, the anisotropy became $\anisotropy_2$. Indeed, this is very plausible if these anisotropies are kept small. Since we get $b\approx 0$ from the limiting case (\ref{Eq:RatioParameterLimit}), we conclude that the anisotropy left that epoch almost with the same magnitude it entered there. 

Since we have got little information on the vanishing of $b$, we might go deeper into its analysis and make it from the scratch. So, setting $\alpha =\pi/6$, for the sake of simplicity, and using the integral (\ref{Eq:RatioParameterIntegral}), we get
\begin{equation}\label{Eq:RatioParameterRadiation}
b = \frac{\anisotropy_1\anisotropy_2\ln(\anisotropy_2/\anisotropy_1)}{\anisotropy_2 - \anisotropy_1 +\Delta}\, ,
\end{equation}
where
\begin{equation}
\Delta = \anisotropy_1\anisotropy_2\ln\sqrt{\left(\frac{1+\anisotropy_2}{1+\anisotropy_1}\right)\left(\frac{1-\anisotropy_1}{1-\anisotropy_2}\right)}\, .
\end{equation}
As we let the universe to expand $N$ e-folds during this epoch, $\ln(a_2/a_1)=N$, we get $\ln(\anisotropy_2/\anisotropy_1)=N b$. Hence, as we  note that $\anisotropy_1 < \anisotropy_2$ implies $\Delta >0$, we obtain
\begin{equation}
\frac{\anisotropy_2-\anisotropy_1}{\anisotropy_1} \le \frac{\anisotropy_2-\anisotropy_1}{\anisotropy_1\anisotropy_2}  < N\, ,
\end{equation}
that is,
\begin{equation}
\anisotropy_1 < \anisotropy_2< (N+1)\anisotropy_1\, ,
\end{equation}
which means that $b < [\ln(N+1)]/N$. In other words, the change in the anisotropy magnitude along the radiation era, as the universe expands $N$ e-folds, is not greater than $\anisotropy_2 \approx (N+1)\anisotropy_1$.

\subsubsection{From radiation-like to stiff-matter-like periods}

In the isotropic model, the interval $1/3 < \state \le 1$ is a candidate for the epoch between the inflationary and the radiation eras at ``ultrahigh" densities. It contains the limiting stiff matter condition ($\state =1$), for which the sound propagation attains the speed of light \cite{Zeldovich_1962, Chavanis_2015}. Interesting enough, our model has naturally separated this regime from the others. Indeed, it appears as the physically relevant interval for which the origin is an unstable equilibrium. Hence, the universe tends to anisotropize even for arbitrarily small and non-vanishing values of $\anisotropy$. This is quite unexpected, for the tidal forces are still isotropic.
\begin{figure}[h!]
\centerline{
\includegraphics[width=5cm]{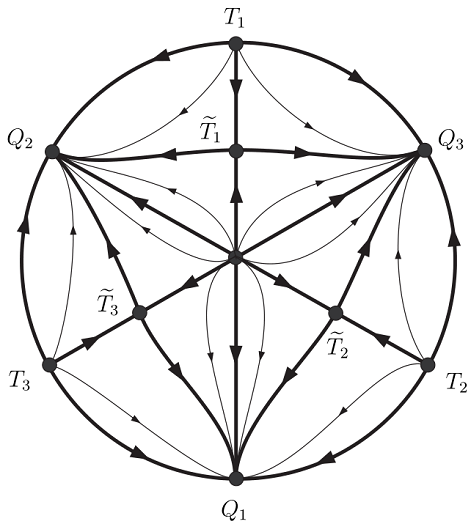}}
\vspace{-0.4cm}
\caption{\small The phase portrait in the Kasner disc ($\anisotropy \le 1$) for $1/3 < \state < 1$. Any solution but the FLRW universe at the center anisotropizes. In the late-time regime they approach the LRS universes at the $Q$-Taub points, with $\anisotropy=1$. The only exceptions are in the lines $\alpha=\pi/2+2n\pi/3$, where $\anisotropy \to (3\state -1)/2$ as $a \to \infty$, the $\widetilde{T}$'s saddle-type equilibrium points. }
\label{Fig4}
\end{figure}

\begin{figure}[htp!]
\centerline{
\includegraphics[width=5cm]{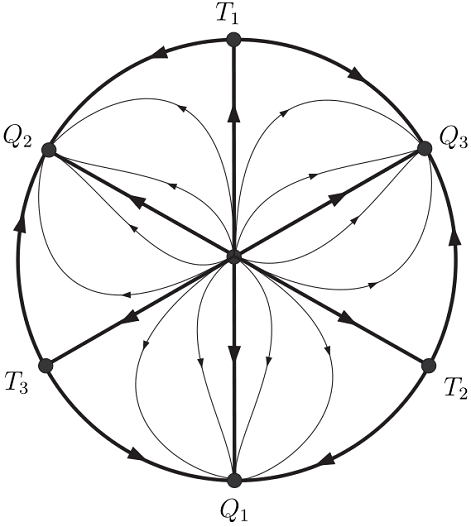}}
\vspace{-0.4cm}
\caption{\small The phase portrait in the Kasner disc ($\anisotropy \le 1$) for $\state = 1$. Any solution but the FLRW universe at the center anisotropizes to the borders of the Kasner disc ($\anisotropy \to 1$ as $a \to \infty$). The saddle-type equilibrium points ($\widetilde{T}$'s) have coalesced with the $T$-Taub points at $\anisotropy=1$. }
\label{Fig5}
\end{figure}

As we allow the equation of state to run from $\state = 1/3$ to $\state =1$, the saddle-type equilibrium points $\widetilde{T}$'s appear at $\anisotropy=(3\state-1)/2$ and $\alpha=\pi/2 +2(1-k)\pi/3$. In the late-time regime, the solutions approach the LRS universes at the $Q$-Taub points, with $\anisotropy=1$, except the isotropic model at the origin and those on the straight line connecting it to $\widetilde{T}$'s, for which $\anisotropy \to (3\state -1)/2$ as $a \to \infty$ (see  Fig. \ref{Fig4}).

The phase portrait in the Kasner disc for the stiff-matter-like situation is plotted in Fig. \ref{Fig5}. In this case, the inner equilibrium points $\widetilde{T}$'s have coalesced with the $T$-Taub points, so that any solution but the isotropic FLRW model at $\anisotropy=0$ tends to the maximum anisotropy $\anisotropy \to 1$ at late-times.

From the formula (\ref{Eq:RatioParameterLimit}), we get $b \approx (3\state-1)/2$. This means that in the stiff-matter-like era, if there has been one, with $\state=1$ and $b \approx 1$,  have the universe passed through $N$ e-folds during this high-density period, the anisotropy would increase just as well.

\section{Final remarks}

In this manuscript, we have analyzed the spatially flat spacetimes under adiabatic expansion and isotropic tidal forces. The total pressure has naturally been divided into the thermodynamical ($P_T$) and bulk ($P_B$) components, the first emulating its isotropic counterpart while the other exists due to anisotropic effects. We analyzed the dynamics of the Hubble anisotropy in the Kasner disc during the different epochs when the ratio $\state =\pressure_T/\energy$ could be held constant. The final framework is a simple and physically relevant scenario where the anisotropy can be understood in its fully non-linear aspects, which was dully depicted in the Kasner disc. For instance, the reader can have a glimpse of the overall aspects of the system and how it changes with $\state$ just by passing from the figures  \ref{Fig1} to \ref{Fig5}, from the isotropizing environment with $\state = -1$ (Fig. \ref{Fig1}) to the completely anisotropizing dynamics with $\state = 1$ (Fig. \ref{Fig5}). 

For the very early universe, our model has much more to tell about the inflationary than the BKL scenario. The reason for this is simple: while the first is highly compatible with the hypothesis of isotropic tidal forces, the chaotic behavior of the second would hardly let this characteristic to be attained. Hence, as we assume inflation took place, the isotropic components of the energy-momentum tensor for the scalar field driving it would overcome the energy density and thermodynamical pressure, so that $\pressure_T \approx - \energy$. The universe would commence a strong isotropization process, with $\anisotropy$ decreasing twice as fast as its expansion rate ($b \approx 2$). At the end of this era, when $\state \sim -1/3$ \cite{Zhou2022}, the (pre)reheating epoch would have begun, with $-1/3 \le \state \le 1$. As we put the e-fold duration of inflation and reheating as $N_{inf}$ and $N_{rh}$, with $\state =-1$ and $\state = 1$, respectively, we would have a estimate for the net decrease of the anisotropy as $e^{-N}$, where $N \sim 2 N_{inf} - N_{rh}$, according to our analysis following the formula (\ref{Eq:RatioParameterLimit}). Recent estimates point to $N_{inf} \sim N_{rh} \sim 60$ \cite{ellis_mac_marteens,Zhou2022}. This would led the anisotropy still insignificant to be detected in the late-time sky, for the periods coming afterwards which are dominated by radiation (anisotropization with $b \approx 0$), dust (anisotropization with $b \approx 1/2$) and dark energy (isotropization with $b \approx 2$), would not last long enough to significantly change this tiny scale. Notwithstanding, we see that a possible detection of anisotropy in the Hubble sky \cite{Schucker_2014, Zhao_2022} is not necessarily in contradiction with inflation, but instead, might shed some light in the period just following it. In principle, our model could favor a quiescent era \cite{Barrow_1978} lasting about twice the preceding inflationary one. Needless to say, this statement is of speculative character. Despite that, this is an example of how our findings can drive us to a qualitative scenario that could hardly be achieved from the perturbative approach, at least in their completeness.      

Regardless of the estimates for the very early universe, our model was designed to attend a gap in the cosmology literature. Whether or not the small perturbations around isotropic models are enough to describe our cosmic observations \cite{Buchert_2015,green2015comments} is of less importance here. Our aim resides in the belief that a comprehensive knowledge of our universe could be achieved only as we understand Einstein's equations and the concepts brought by them as a whole. In that direction, we have presented a model that could be considered of physical interest just as a perfect fluid can \footnote{Anisotropy is often considered in conjunction to perfect fluids, for which $\Pi_{ij}=0$. Typically, such models have a simple dynamics, with $\dot{\alpha}=0$ and $\sigma_{ij} \sim a^{-3}$.}, but containing much more of the complex aspects of the anisotropies than the latter. In particular, we have a broader view of their global dynamics with the aid of the Kasner disc as well as we can follow their tracks along the different epochs of the cosmic expansion, just as we did with the Hubble anisotropy by using the parameter $b$, which enhanced our view of this furtive concept. Last but not least important, the constraints our model imposes on the observational data, in the lines of Ref. \cite{LGGomes_2021}, is something we have not touched upon in this manuscript. We leave this issue to be accomplished elsewhere.

\section*{Acknowledgments}
FSD is partially supported by Funda\c c\~ao de Amparo \`a Pesquisa do Estado de Minas Gerais -- FAPEMIG [grant number APQ--01158--17].
LFM is partially supported by Funda\c c\~ao de Amparo \`a Pesquisa do Estado de Minas Gerais -- FAPEMIG [grant number APQ--01105--18] and by Conselho Nacional de Desenvolvimento Cient\'ifico e Tecnol\'ogico -- CNPq [grant number 311921/2020--5].

\bibliography{ref}

\begin{thebibliography}{10}

\bibitem{Soltis_2019}
John Soltis, Arya Farahi, Dragan Huterer, and C.~Michael Liberato.
\newblock Percent-level test of isotropic expansion using type ia supernovae.
\newblock {\em Phys. Rev. Lett.}, 122:091301, Mar 2019.

\bibitem{Tedesco_2019}
\"Ozg\"ur Akarsu, Suresh Kumar, Shivani Sharma, and Luigi Tedesco.
\newblock {Constraints on a Bianchi type I spacetime extension of the standard
  $\Lambda$CDM model}.
\newblock {\em Phys. Rev. D}, 100(2):023532, 2019.

\bibitem{Saadeh_PRL}
Daniela Saadeh, Stephen~M. Feeney, Andrew Pontzen, Hiranya~V. Peiris, and
  Jason~D. McEwen.
\newblock How isotropic is the universe?
\newblock {\em Phys. Rev. Lett.}, 117:131302, Sep 2016.

\bibitem{Isotropy_CMB_2020}
{Planck Collaboration}, {Akrami, Y.}, {Ashdown, M.}, {Aumont, J.},
  {Baccigalupi, C.}, {Ballardini, M.}, {Banday, A. J.}, {Barreiro, R. B.},
  {Bartolo, N.}, {Basak, S.}, {Benabed, K.}, {Bersanelli, M.}, {Bielewicz, P.},
  {Bock, J. J.}, {Bond, J. R.}, {Borrill, J.}, {Bouchet, F. R.}, {Boulanger,
  F.}, {Bucher, M.}, {Burigana, C.}, {Butler, R. C.}, {Calabrese, E.},
  {Cardoso, J.-F.}, {Casaponsa, B.}, {Chiang, H. C.}, {Colombo, L. P. L.},
  {Combet, C.}, {Contreras, D.}, {Crill, B. P.}, {de Bernardis, P.}, {de Zotti,
  G.}, {Delabrouille, J.}, {Delouis, J.-M.}, {Di Valentino, E.}, {Diego, J.
  M.}, {Dor\'e, O.}, {Douspis, M.}, {Ducout, A.}, {Dupac, X.}, {Efstathiou,
  G.}, {Elsner, F.}, {En\ss{}lin, T. A.}, {Eriksen, H. K.}, {Fantaye, Y.},
  {Fernandez-Cobos, R.}, {Finelli, F.}, {Frailis, M.}, {Fraisse, A. A.},
  {Franceschi, E.}, {Frolov, A.}, {Galeotta, S.}, {Galli, S.}, {Ganga, K.},
  {G\'enova-Santos, R. T.}, {Gerbino, M.}, {Ghosh, T.}, {Gonz\'alez-Nuevo, J.},
  {G\'orski, K. M.}, {Gruppuso, A.}, {Gudmundsson, J. E.}, {Hamann, J.},
  {Handley, W.}, {Hansen, F. K.}, {Herranz, D.}, {Hivon, E.}, {Huang, Z.},
  {Jaffe, A. H.}, {Jones, W. C.}, {Keih\"anen, E.}, {Keskitalo, R.}, {Kiiveri,
  K.}, {Kim, J.}, {Krachmalnicoff, N.}, {Kunz, M.}, {Kurki-Suonio, H.},
  {Lagache, G.}, {Lamarre, J.-M.}, {Lasenby, A.}, {Lattanzi, M.}, {Lawrence, C.
  R.}, {Le Jeune, M.}, {Levrier, F.}, {Liguori, M.}, {Lilje, P. B.}, {Lindholm,
  V.}, {L\'opez-Caniego, M.}, {Ma, Y.-Z.}, {Mac\'{\i}as-P\'erez, J. F.},
  {Maggio, G.}, {Maino, D.}, {Mandolesi, N.}, {Mangilli, A.},
  {Marcos-Caballero, A.}, {Maris, M.}, {Martin, P. G.},
  {Mart\'{\i}nez-Gonz\'alez, E.}, {Matarrese, S.}, {Mauri, N.}, {McEwen, J.
  D.}, {Meinhold, P. R.}, {Mennella, A.}, {Migliaccio, M.},
  {Miville-Desch\^enes, M.-A.}, {Molinari, D.}, {Moneti, A.}, {Montier, L.},
  {Morgante, G.}, {Moss, A.}, {Natoli, P.}, {Pagano, L.}, {Paoletti, D.},
  {Partridge, B.}, {Perrotta, F.}, {Pettorino, V.}, {Piacentini, F.}, {Polenta,
  G.}, {Puget, J.-L.}, {Rachen, J. P.}, {Reinecke, M.}, {Remazeilles, M.},
  {Renzi, A.}, {Rocha, G.}, {Rosset, C.}, {Roudier, G.}, {Rubi\~no-Mart\'{\i}n,
  J. A.}, {Ruiz-Granados, B.}, {Salvati, L.}, {Savelainen, M.}, {Scott, D.},
  {Shellard, E. P. S.}, {Sirignano, C.}, {Sunyaev, R.}, {Suur-Uski, A.-S.},
  {Tauber, J. A.}, {Tavagnacco, D.}, {Tenti, M.}, {Toffolatti, L.}, {Tomasi,
  M.}, {Trombetti, T.}, {Valenziano, L.}, {Valiviita, J.}, {Van Tent, B.},
  {Vielva, P.}, {Villa, F.}, {Vittorio, N.}, {Wandelt, B. D.}, {Wehus, I. K.},
  {Zacchei, A.}, {Zibin, J. P.}, and {Zonca, A.}
\newblock Planck 2018 results - vii. isotropy and statistics of the cmb.
\newblock {\em A\&A}, 641:A7, 2020.

\bibitem{Bengaly_2019}
Carlos~A.P. Bengaly, Roy Maartens, Nandrianina Randriamiarinarivo, and Albert
  Baloyi.
\newblock Testing the cosmological principle in the radio sky.
\newblock {\em Journal of Cosmology and Astroparticle Physics},
  2019(09):025--025, sep 2019.

\bibitem{Weinberg_2008}
S.~Weinberg.
\newblock {\em Cosmology}.
\newblock Cosmology. OUP Oxford, 2008.

\bibitem{BKL82}
V.A. Belinskii, I.M. Khalatnikov, and E.M. Lifshitz.
\newblock A general solution of the einstein equations with a time singularity.
\newblock {\em Advances in Physics}, 31(6):639--667, 1982.

\bibitem{Misner67}
Charles~W. Misner.
\newblock Neutrino viscosity and the isotropy of primordial blackbody
  radiation.
\newblock {\em Phys. Rev. Lett.}, 19:533--535, Aug 1967.

\bibitem{Misner68}
Charles~W. {Misner}.
\newblock {The Isotropy of the Universe}.
\newblock {\em \apj}, 151:431, February 1968.

\bibitem{CollinsHawking1973}
C.~B. {Collins} and S.~W. {Hawking}.
\newblock {Why is the Universe Isotropic?}
\newblock {\em \apj}, 180:317--334, March 1973.

\bibitem{Barrow82}
J.~D. {Barrow}.
\newblock {The Isotropy of the Universe}.
\newblock {\em QJRAS}, 23:344, September 1982.

\bibitem{Berger2014}
Beverly~K. Berger.
\newblock {\em Singularities in Cosmological Spacetimes}, pages 437--460.
\newblock Springer Berlin Heidelberg, Berlin, Heidelberg, 2014.

\bibitem{LeBlanc_1997}
Victor~G LeBlanc.
\newblock Asymptotic states of magnetic bianchi i cosmologies.
\newblock {\em Classical and Quantum Gravity}, 14(8):2281--2301, aug 1997.

\bibitem{Calogero2008}
Simone Calogero and J.~Mark Heinzle.
\newblock {Dynamics of Bianchi type I solutions of the Einstein equations with
  anisotropic matter}.
\newblock {\em Annales Henri Poincare}, 10:225--274, 2009.

\bibitem{BSGC_2021}
Bruno~B. Bizarria, Gabriel A.~Souza Silva, Leandro~G. Gomes, and William~O.
  Clavijo.
\newblock The oscillatory anisotropy in the spatially flat cosmological models.
\newblock {\em Annals of Physics}, 432:168571, 2021.

\bibitem{DSGM_2022}
Fabio~Scalco Dias, Grasiele~B. Santos, Leandro~G. Gomes, and Luis~Fernando
  Mello.
\newblock {The power-law dependence between the matter-radiation and Hubble
  anisotropies}.
\newblock {\em International Journal of Modern Physics D}, 31(07):2250049, 05
  2022.

\bibitem{Wald1983}
Robert~M. Wald.
\newblock {Asymptotic behavior of homogeneous cosmological models in the
  presence of a positive cosmological constant}.
\newblock {\em Phys. Rev. D}, 28:2118--2120, 1983.

\bibitem{Schucker_2014}
Thomas Schucker, Andr\'e Tilquin, and Galliano Valent.
\newblock {Bianchi I meets the Hubble diagram}.
\newblock {\em Mon. Not. Roy. Astron. Soc.}, 444(3):2820--2836, 2014.

\bibitem{Zhao_2022}
Dong {Zhao} and Jun-Qing {Xia}.
\newblock {Testing cosmic anisotropy with the E$_{p}$-E$_{iso}$ ('Amati')
  correlation of GRBs}.
\newblock {\em Mon. Not. Roy. Astron. Soc.}, 511(4):5661--5671, April 2022.

\bibitem{MTW}
Charles Misner, Kip~S. Thorne, and John~Archibald Wheeler.
\newblock {\em Gravitation}.
\newblock Princeton University Press, 2017.

\bibitem{ellis_mac_marteens}
G.~F.~R. Ellis, R.~Maartens, and M.~A.~H. MacCallum.
\newblock {\em Relativistic Cosmology}.
\newblock Cambridge University Press, 2012.

\bibitem{Note1}
For the Taub points, see \cite {BSGC_2021}. Any point with Kasner angular
  coordinates $n\pi /6$, $n$ odd, is Locally Rotationally Symmetric (LRS),
  which refers to the more symmetric configuration of the spacetime. A good
  analogy is to compare the ellipsoids of revolution in the Euclidean spaces
  (LRS) with their less symmetric partners.

\bibitem{DLA_2007}
Freddy Dumortier, Jaume Llibre, and Joan Artés.
\newblock {\em Qualitative Theory Of Planar Differential Systems}.
\newblock Springer Verlag, New York, 2007.

\bibitem{Note2}
Indeed, the strong energy condition demands $\protect \ensuremath {\protect
  \mathrm {\rho }}+ 3 \protect \ensuremath {\protect \mathrm {P}}>0$ and
  $\protect \ensuremath {\protect \mathrm {\rho }}+ \protect \ensuremath
  {\protect \mathrm {P}}+ \Pi _i >0$, $i=1,2,3$, $\Pi _i$ the eigenvalues of
  $\Pi _{ij}$. The name refers to the state at which it begins to be violated,
  at least for small anisotropy ($\protect \raisebox {0.01em}{\scalebox
  {0.9}[1.3]{${\scriptstyle \Sigma }$}}\approx 0$).

\bibitem{Zeldovich_1962}
Zel'dovich Ya.~B.
\newblock {The equation of state at ultrahigh densities and its relativistic
  limitations}.
\newblock {\em Journal of Experimental and Theoretical Physics}, 14(05):1143,
  05 1962.

\bibitem{Chavanis_2015}
Pierre-Henri Chavanis.
\newblock Cosmology with a stiff matter era.
\newblock {\em Phys. Rev. D}, 92:103004, Nov 2015.

\bibitem{Zhou2022}
Hua Zhou, Qing Yu, Yu~Pan, Ruiyu Zhou, and Wei Cheng.
\newblock {Reheating constraints on modified single-field natural inflation
  models}.
\newblock {\em Eur. Phys. J. C}, 82(7):588, 2022.

\bibitem{Barrow_1978}
John~D. Barrow.
\newblock {Quiescent cosmology}.
\newblock {\em Nature}, 272:211, 1978.

\bibitem{Buchert_2015}
T~Buchert, M~Carfora, G~F~R Ellis, E~W Kolb, M~A~H MacCallum, J~J Ostrowski,
  S~Räsänen, B~F Roukema, L~Andersson, A~A Coley, and D~L Wiltshire.
\newblock Is there proof that backreaction of inhomogeneities is irrelevant in
  cosmology?
\newblock {\em Classical and Quantum Gravity}, 32(21):215021, oct 2015.

\bibitem{green2015comments}
Stephen~R. Green and Robert~M. Wald.
\newblock Comments on backreaction, 2015.

\bibitem{Note3}
Anisotropy is often considered in conjunction to perfect fluids, for which $\Pi
  _{ij}=0$. Typically, such models have a simple dynamics, with $\protect \dot
  {\alpha }=0$ and $\sigma _{ij} \sim a^{-3}$.

\bibitem{LGGomes_2021}
Leandro~G. Gomes.
\newblock {The nonlinear patterns of the cosmic anisotropy: the spatially flat
  perfect fluid universes}.
\newblock {\em Class. Quant. Grav.}, 39(2):027001, 2022.

\end{thebibliography}

\end{document}